\documentclass[default,iicol,referee,10pt]{sn-jnl}

\jyear{2022}%

\theoremstyle{thmstyletwo}%

\theoremstyle{thmstylethree}%
\newtheorem{definition}{Definition}%

\begin{document}

\title[Article Title]{An Adaptive Channel Reservation MAC Protocol Based on Forwarding Traffic of Key Nodes}


\author[1]{\fnm{Ze} \sur{Liu}}\email{2021202099@mail.nwpu.edu.cn}

\author[1]{\fnm{Bo} \sur{Li}}\email{libo.npu@nwpu.edu.cn}

\author*[1]{\fnm{Mao} \sur{Yang}}\email{yangmao@nwpu.edu.cn}

\author[1]{\fnm{ZhongJiang} \sur{Yan}}\email{zhjyan@nwpu.edu.cn}

\affil*[1]{\orgdiv{School of Electronics and Information}, \orgname{Northwestern Polytechnical University}, \city{Xi'an}, \postcode{710072}, \state{Shaanxi}, \country{P. R. China}}

\abstract{Ad Hoc networks with multi-hop topology are widely used in military and civilian applications. One challenge for Ad Hoc networks is to design efficient Media Access Control (MAC) protocols to ensure the quality of service (QoS). In Ad Hoc networks, there is a kind of node called key node, which undertakes more forwarding traffic than other surrounding nodes. The number of neighbor nodes around key nodes is often large, and the surrounding channel environment and interference are often more complex. Thus, the key nodes can hardly get enough channel access opportunities, resulting in poor end-to-end performance. Therefore, we propose an adaptive channel reservation MAC protocol based on forwarding traffic of key nodes, which is aimed at alleviating the congestion for key nodes. Nodes initiate reservations for future transmission time according to the buffer status before sending packets and then calculate the Weight of Reservation Ability (WRA). The node adaptively adjusts its reservation opportunity by comparing the WRA with neighbor nodes, thus improving the channel access efficiency and ensuring the transmission opportunity of key nodes. Extensive simulation confirms that our proposed FTKN-CRM provides significant improvements in end-to-end performance over the IEEE 802.11ax protocol and other reservation access protocols.}




\maketitle

\begin{figure*}[htbp]
    \centering
    \includegraphics[width=1\textwidth]{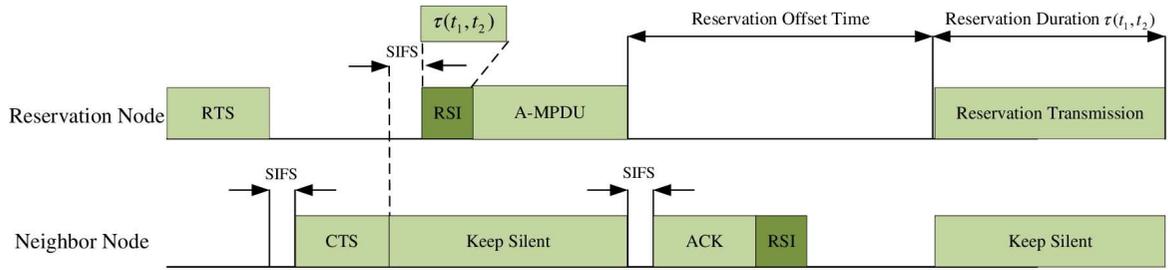}
    \caption{Basic reservation channel access mechanism.}
	\label{fig1}
\end{figure*}

\section{Introduction}\label{Introduction}
In wireless communications, the Ad Hoc network is a kind of wireless network without a central node and is formed by multiple nodes autonomously. Because its self-organizing characteristic enables rapid deployment, it is widely used in the situations such as emergency and disaster relief scenarios \cite{1}.

In wireless Ad Hoc networks, media access control (MAC) protocols based on carrier sense multiple access with collision avoid (CSMA/CA) mechanism are widely used due to the advantages of flexibility and decentralized resource allocation. However, as the traffic loads increase in the multi-hop network scenario, the end-to-end performance degrades significantly due to conflicts arising from random contention and the phenomenon of hidden terminals, resulting in unsatisfactory quality of services (QoS) for real-time traffic \cite{2}. In the channel reservation mechanism, when a node completes a channel transmission opportunity, if there is still real-time traffic to be transmitted in the queue, it can carry the reservation information in the frame to reserve channel access in advance for a certain period of time in the future. Neighbor nodes will remain silent at the corresponding time after receiving it, thus reducing the disorderly competition for channel resources. Therefore, introducing channel reservation under CSMA/CA mechanism is an efficient way to guarantee the QoS in Ad Hoc networks.

There are several existing works on channel reservation. Li \emph{et al.} \cite{3} proposes a multi-step distributed reservation mechanism, where each node maintains counters corresponding to the time slots when multiple subsequent packets are ready to be sent, and multiple broadcasts enable neighbor nodes to efficiently obtain reservation information and thus avoid conflicts. Yang \emph{et al.} \cite{4} tries to improve the throughput by performing scheduled transmission of packets on the data channel under a multi-channel MAC protocol, which results in a significant reduction in the traffic load on the control channel. Xia \emph{et al.} \cite{5} proposes a reservation MAC mechanism based on the reservation of future time slots, according to link transmission delays in the hydroacoustic channel. Cheng \emph{et al.} \cite{6} proposes to improve channel utilization by canceling the reservation if no packets are available at the reservation time.

However, it can be seen that the existing reservation mechanism is usually for fully connected network, but there are problems with the existing reservation mechanism in multi-hop topology because there are some key nodes in Ad Hoc networks that have more complex surrounding topology and take more relay-traffic, and such nodes play a more important role in multi-hop transmission due to the larger amount of relay-traffic, but in the traditional channel reservation process these key nodes get less reservation opportunities due to more competing neighbor nodes, which makes the end-to-end performance degrade.

In this paper, we propose an adaptive channel reservation MAC protocol based on forwarding traffic of key nodes named FTKN-CRM that can improve the end-to-end performance during multi-hop transmission. For FTKN-CRM, the nodes compare their traffic type and queue utilization with neighbor nodes when initiating channel reservations, and adaptively adjust the reservation parameters to increase the transmission opportunities of key nodes, thus improving the end-to-end performance of the network.

The remainder of this paper is organized as follows. Section~\ref{Problem Analysis} describes the causes of congestion problems and the necessity of improving the transmission opportunities for key nodes through MAC protocols. Section~\ref{System Model} presents the system model, including the definition and calculation methods of reservation parameters. Section~\ref{PROPOSED FTKN-CRM PROTOCOL} describes the proposed protocol and frame structure. In Section~\ref{SIMULATION RESULTS} the simulation results and analysis are presented, and Section~\ref{Conclusion} provides the conclusions.

\section{Problem Analysis}
\label{Problem Analysis}
\subsection{Basic Channel Reservation Mechanism}
\label{Basic Channel Reservation Mechanism}
Figure~\ref{fig1} shows the channel access model based on the channel reservation mechanism. The node first accesses the channel through competition. In the transmission period, if there is extra real-time traffic in the queue, the node calculates its required transmission duration based on the amount of real-time traffic to be transmitted as channel reservation duration. Then it selects the period time that does not conflict with other nodes as its reservation period, which need to meet the minimum reservation offset time. Finally, the above channel reservation information is aggregated in the AMPDU of this transmission through the Reservation Instruction MAC Protocol Data Unit (RSI MPDU). The receiving node forwards the RSI MPDU in the ACK to overcome the hidden terminal. When neighbor nodes receive the RSI MPDU, they will keep silent, and goes into sleep mode at the corresponding reservation time. Where the reservation offset time is defined as the time interval between the reservation time and the end of this A-MPDU transmission, and the reservation duration is defined as the duration of the reservation period.
\subsection{Analysis of Multi-hop Network Congestion Problem}
\label{Analysis of Multi-hop Network Congestion Problem}
In Ad Hoc networks, the nodes are connected in a multi-hop transmission via wireless links. Since there is no fixed infrastructure, when nodes communicate with each other, any relay node may become congested due to the full buffer
, which requires MAC protocols to ensure good QoS in this situation \cite{7}.

\begin{figure}[hb]
    \centering
    \includegraphics[width=0.3\textwidth]{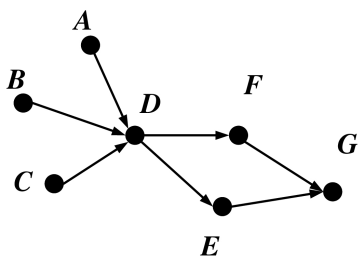}
    \caption{Multi-hop network congestion problem.}
	\label{fig2}
\end{figure}
Consider the simple topology shown in Figure~\ref{fig2}, in which both nodes send services through relay nodes for forwarding traffic, and most of the services converge at the queue of node D. The more nodes competing with node D makes its transmission opportunities less and cannot meet the demand for more relay-traffic. When the traffic loads are heavier, it is easy for the neighbor nodes to continuously reserve the channel and make it difficult for node D to get the channel reservation opportunity. To address the above problems,  much literature has proposed some solutions, and these approaches can be classified into three categories according to the protocol layer: TCP congestion control, routing path selection, and MAC layer approaches.

End-to-end TCP congestion control algorithms based on congestion windows can be applied to communication protocols, but Kittali  \emph{et al.} \cite{8} points out that congestion avoidance techniques based on TCP protocols are not suitable for Ad Hoc networks, and TCP congestion mechanisms are mainly used to solve wired congestion problems and cannot handle wireless congestion well. In the research for routing protocols, some literature mentions improved solutions for congestion control. Monisha \emph{et al.} \cite{9} proposes methods to avoid congestion based on network traffic estimation and path splitting by shifting traffic to alternative paths or dropping it along optional paths thus reducing congestion.

However, Jin \emph{et al.} \cite{11}  points out that congestion avoidance is not considered in most of the current routing protocols. Since traffic is usually shifted to less congested paths by the traffic estimation algorithm, which relies on the accuracy of the estimation and requires high real-time performance, the traffic estimation algorithm can not predict congested links quickly and accurately when facing bursty services \cite{12}. Since there are differences in the quality of different links, a single node in a routing algorithm usually chooses the path with better link quality to expect better performance for itself, and if many nodes choose the better path it will cause congestion in the relay nodes on the better path, while deliberately avoiding the path with higher traffic link quality will reduce the performance of this node \cite{13}.

It is more efficient and necessary to perform congestion control at the MAC layer because the channel state is directly accessible. Chou \emph{et al.} \cite{14} proposes a mechanism to calculate the contention window according to the total hops and the remaining hops to reduce data conflicts by optimizing the contention window in the transmission path, but does not take into account the inter-node variability. Ibukunoluwa \emph{et al.} \cite{15} proposes to dynamically change the size of the competition window according to the number of active nodes in the network to mitigate conflicts in congested areas in Ad Hoc networks, but the relative importance relationship with neighboring nodes is not considered, and the problem of increased conflicts exists when there are multiple key nodes around all using lower competition windows. Reddy \emph{et al.} \cite{15.1} solves congestion and scheduling problems according to a cross-layer design, using the successive interference cancelation (SIC) model. Chao \emph{et al.} \cite{16} pointed out that if the MAC protocol is distributed equally to each node, without considering the aggravation of the relay-traffic load close to the gateway nodes, it will make the end-to-end throughput degraded. Therefore it assigns higher priority to the relay traffic transmission for the nodes close to the gateway. However, this scheme is based on TDMA with a central control node, which is not convenient and flexible enough to meet the latency demand of delay-sensitive real-time traffic.

Therefore, although many existing MAC schemes consider the conflict problem in congested areas, there are fewer MAC schemes that notice the importance of nodes in multi-hop transmission. However, in order to increase the priority of the important nodes, these schemes only adjust transmission parameters like the contention window based on their own congestion information, such as the number of packet hops or the location, and compare to a fixed threshold value, without considering the traffic load of neighbor nodes. When there are many busy key nodes at the same time, the adjustments may instead exacerbate the conflict and degrades network performance.To address the above  problem, this paper proposes an adaptive channel reservation protocol to alleviate the phenomenon.

\section{System Model}
\label{System Model}
\subsection{Network Topology}
The network topology in this paper is a multi-hop wireless network. When a node sends data to another node, it may need to be forwarded through many relay nodes. The routing path is specified by the network layer protocol. Traffic that reachs the MAC layer includes the following two types: Traffic generated by this node is defined as local-traffic. Traffic generated by other nodes that need to be forwarded by this node are defined as relay-traffic. It is assumed that all nodes in the network adopt the same channel frequency band and access the channel through the channel reservation mechanism based on CSMA/CA mechanism extension.

\subsection{Definition of system model}
\label{Definition of system model}
\begin{definition}\label{definition1}
Relay-Traffic and Local-Traffic.

The services generated by this node in the network are defined as local-traffic, and the services generated by other nodes that need to be forwarded by this node are defined as relay-traffic.
\end{definition}
\begin{definition}\label{definition2}
Weight of Reservation Ability.

The constructed parameters that reflect the importance of the node in the multi-hop transmission, calculated according to the relay-traffic and local-traffic, are denoted by WRA.
\end{definition}
\begin{definition}\label{definition3}
Contention Range.

In the channel reservation mechanism, the area where the nodes involved in the competition for channel resources with this node is defined as the competition range, which is denoted as $R_c$. The set of all nodes within the competition range is defined as the set of competing nodes, which is denoted as $S_c$. The set consisting of N hop neighbors is denoted as $S_H^N$. Due to the mechanism of RTS/CTS and RSI MPDUs forwarded by ACK, the channel reservation information of each node is submitted to its second-hop neighbor nodes, so the $S_c$ of each node in the Ad Hoc network is the set of all first-hop and second-hop neighbor nodes. Therefore we conclude that $S_c=S_H^1\cup S_H^2$.
\end{definition}
\begin{definition}\label{definition4}
Basic Reservation Offset Time.

Without considering the inter-node difference, the minimum reservation offset time that can guarantee that all nodes in the $S_c$ can get transmission opportunities is defined as the basic reservation offset time, denoted by $T_{basic}$.
\end{definition}
\begin{definition}\label{definition5}
Reservation Capacity Correction Factor.

The ratio of this node’s WRA to the average WRA of neighbor nodes ($N_i\in S_c$) is defined as the reservation capacity correction factor, denoted by $\bar{\lambda}$. This dimensionless quantity reflects the importance of this node relative to the $N_i$ in $R_c$ for services forwarding.
\end{definition}
\begin{definition}\label{definition6}
Correction Reservation Offset Time.

The basic reservation offset time corrected based on the $\bar{\lambda}$ is denoted by $T_{offset}$, which is configured with different reservation priorities according to the importance of this node in multi-hop transmission.
\end{definition}
\begin{definition}\label{definition7}
N-step Inertia Factor.

The ability to make $T_{offset}$  maintain the same value as the previous n reservations is defined as the n-step inertia factor, denoted by $w_{n}$ . Since frequent changes in the WRA of neighbor nodes may lead to oscillations in each
reservation offset time, which is not conducive to the convergence of the network to a stable state. We define the correction reservation offset time calculated at the m-th reservation is $ T_{offset}^m $. Therefore, the $T_{offset}^{n+1}$ of each calculation can be corrected according to the $T_{offset}^{i}(i\in[1,n])$ of the previous n reservations,
and $w_{n}$ denotes the impact of the previous n $T_{offset}^{i}$ on the calculation of $T_{offset}^{n+1} $ .
\end{definition}
\begin{definition}\label{definition8}
Congestion Queue Utilization Ratio.

If $U_{buffer}>Th_{buffer}$, we consider the node is in congestion status, where the $U_{buffer}$ means the ratio of the total number of packages in the buffer to the max size and the $Th_{buffer}$ is the corresponding threshold.
\end{definition}
\begin{definition}\label{definition9}
Transmission Time Fragmentation.

The time required for the entire process of obtaining a transmission opportunity under the CSMA/CA mechanism is denoted by $T_{min }$ and the interval time between adjacent reservation duration is denoted by $T_{bet}$.
 If $T_{bet }<T_{min }$, this interval cannot be fully utilized.
The phenomenon that generates the above channel resources wasted is defined as transmission time fragmentation, and the corresponding transmission time is denoted by $T_{frag}$.
\end{definition}
\begin{definition}\label{definition10}
Maximum Reservation Ratio.

The maximum proportion of channel resources allowed to be occupied by reservation in the network is defined as $Th_{res} (Th_{res}\in[0,1])$. To avoid the channel resources completely occupied by reservation period time, which results in transmission opportunities not available to nodes with non-real-time traffic, the basic reservation offset time is extended to $T_{offset}/Th_{res}$ to ensure that some slots in the reservation cycle are reserved for random contention.
\end{definition}
\subsection{Calculation of WRA}
\label{Calculation of WRA}
Marchang \emph{et al.} \cite{17} points out that packets that have traveled more hops in a longer topology need higher priority transmission opportunities, so nodes with more relay-traffic should get more reservation opportunities. Construct the WRA that can reflect the reservation priority and calculate it as follows:
\begin{equation}
WRA=\sum\nolimits_{m=0}^{N_{HopNum}} \alpha _{m} N_{m}
\label{eq1}
\end{equation}

Where $N_{m}$  denotes the total length of packets in the queue that have been forwarded m times and $\alpha _{m}$ denotes the weight factor for packets that have been forwarded m times. The weight factor of local-traffic is configured smaller than relay-traffic's because the relay-traffic can better reflect the characteristics of key nodes.

\subsection{Calculation of basic reservation offset time}
\label{Calculation of basic reservation offset time}

To ensure fairness, the reservation offset time should satisfy that all nodes in the competitive range have access to data transmission opportunities in a reservation cycle. Therefore the number of packet bytes in the queue of other node can be calculated by the WRA in the recorded contention range and the average weight factor. The transmission time is calculated based on the physical layer transmission rate, which in turn calculates the total transmission time of neighboring nodes in the competitive range, and finally, the reservation cycle is extended according to the $Th_{res}$ to reserve remaining channel resources for random competition. Thus the $T_{basic}$ is calculated as follows.
\begin{equation}
T_{basic}=\frac{\sum\limits_{i=0}^{N_{neighbor}}min(D_{max},\frac{WRA_i}{\bar \alpha} V_{phy}+T_{Access})}{Th_{res}}
\label{eq2}
\end{equation}

Where the maximum reservation duration is denoted by $D_{max}$, the average weight factor is denoted by $\alpha$, the physical layer transmission rate is denoted by $V_{phy}$,  the channel access time and control frame overhead other than packet transmission in the channel access process is denoted by $T_{Access}$, and the number of neighbor nodes in the contention range is denoted by $N_{neighbor}$.
\subsection{Calculation of Correction Reservation Offset Time}
\label{Calculation of Correction Reservation Offset Time}
The calculation formula of $\bar{\lambda }$ is shown as~\eqref{eq3}, where the total number
of nodes within the competition range and the WRA of neighbor node i are denoted
by $n_{neibor}$ and $WRA_{i}$.
\begin{equation}
\bar{\lambda}=\frac{WRA_0}{{\sum\limits_{i=1}^{N_{neighbor}}WRA_i /N_{neighbor}}}
\label{eq3}
\end{equation}

According to the defination of $Th_{buffer}$ and~\eqref{eq1}, the WRA corresponding to congested state can be derived as~\eqref{eq4}. Where the $L_{buffer}$ represents the average buffer length.
\begin{equation}
WRA_{conges}=\bar\alpha Th_{buffer}L_{buffer}
\label{eq4}
\end{equation}

$WSA_{max}$ represents the recorded maximum WSA of neighbor nodes in the contention range, and the minimum reservation offset time in the non-congested state is denoted by $T_{min}$. By comparing the current $WSA_{max}$ with the threshold of the congestion state, we can determine whether there are neighbor nodes in the congestion state, so the correction reservation offset time calculation formula is given by
\begin{equation}
T_{offset} = \begin{cases}
T_{\min } \quad ,&WRA_{\max } <WRA_{conges}  \\
\frac{T_{basic} }{\bar{\lambda } }\quad ,&WRA_{\max } \geq WRA_{conges}   \\
\end{cases}
\label{eq5}
\end{equation}

From~\eqref{eq3} and~\eqref{eq5}, it can be seen that the WRA of nodes with more relay-traffic is higher than the average WRA of neighbor nodes. Therefore, the Adopted $T_{offset}$ of these nodes is smaller than that of neighbor nodes, which allows key nodes to get more reservation opportunities than ordinary nodes. And since the correction factor is a dimensionless quantity derived from comparison rather than the fixed threshold, the scheme will adopt more conservative reservation parameters when there are multiple nodes with heavy loads in the contention range, thus ensuring that it is widely applicable in single or multiple critical nodes, saturated or unsaturated scenarios.

~\eqref{eq6} shows the de-oscillation of the corrected reservation offset time according to the inertia factor. For the calculated offset time used for this appointment, the percentage of the average appointment offset time calculated in the past n times is $w_n$. Therefore, even if there is a bursty service or other special circumstances, the node reservation offset time will not change drastically due to the limitation of $w_n$, making the network a relatively stable operation.
\begin{equation}
T_{offset}^{i}=\frac{w_{n}}{n}\sum_{m=1}^{n}T_{offset}^{i-m}+(1-w_n)T_{offset}^i,w\in[0,1)
\label{eq6}
\end{equation}

Finally, in order to avoid calculating extreme $T_{offset}$ in special cases, ~\eqref{eq7} specifies the corrected offset time range, where $Th_{low}$ and $Th_{high}$ are the lower and upper threshold for the correction of $Th_{basic}$ by the weight ratio value, respectively.
\begin{equation}
\frac{T_{offset}}{T_{basic}}\in[Th_{low},Th_{high}]
\label{eq7}
\end{equation}
\subsection{Calculation of Maximum fragmentation time interval}
\label{Calculation of Maximum fragmentation time interval}
According to Definition~\ref{definition10}, the maximum fragmentation interval is the transmission time of A-MPDU aggregation of m and the corresponding control frame time, calculated as follows:
\begin{equation}
\begin{aligned}
T_{frag}=T_{RTS}+T_{SIFS}+T_{CTS}+T_{SIFS} \\
+m\times T_{DATA}+T_{SIFS}+T_{ACK}
\end{aligned}
\label{eq8}
\end{equation}
\section{PROPOSED FTKN-CRM PROTOCOL}
\label{PROPOSED FTKN-CRM PROTOCOL}
In this section, the FTKN-CRM protocol proposed in this paper is introduced. The FTKN-CRM protocol not only follows the basic reservation access process, but also includes the following 3 phases: WRA updating phase, reservation phase and channel access phase.The pseudocode of protocol is shown in Algorithm~\ref{algorithm1}.

\begin{algorithm}
\caption{FTKN-CRM}\label{algorithm1}
\begin{algorithmic}[1]
\Require {$\left\{WRA_{0},WRA_{1}...,WRA_{N_{neighbor}}\right\};$ }
\State \noindent\textbf{WRA updating phase:}
\State $WRA_{0}\gets \sum\nolimits_{m=0}^{N_{HopNum}} \alpha _{m} N_{m}$
\State $ \overline{\lambda} \gets \frac{WRA_{0}}{\sum\limits_{i=1}^{N_{neighbor}}WRA_{i}/N_{neighbor}} $
\State \noindent\textbf{Reservation phase:}
\State $T_{basic}\gets \sum\limits_{i=0}^{n_{neighbor}}min(D_{max}, \frac{WRA_{i}}{\overline{\alpha}V_{phy}+T_{Access}}/Th_{res})$
\If {$\overline{\lambda} \textgreater Th_{high}$}
\State {$ \overline{\lambda}\gets Th_{high}$}
\ElsIf {$\overline{\lambda}\textless Th_{low}$}
\State {$ \overline{\lambda}\gets Th_{low}$}
\EndIf
\If {$WRA_{max}\textless WRA_{conges}$}
\State{$T_{offset} \gets T_{min}$}
\Else
\State {$T_{offset} \gets T_{basic}/ \overline{\lambda}$}
\EndIf
\If {$T_{offset}^i-T_{offset}^{last}\textless T_{frag} $}
\State {$ T_{offset}^i \gets T_{betmin}+T_{offset}^{last}$}
\EndIf
\State \noindent\textbf{Channel Access phase:}
\If {$\overline{\lambda} \textgreater Th_{less}$}
\State {$CW \gets CW_{low}$}
\ElsIf {There are no links that require reservations}
\State {$CW \gets CW_{high}$}
\EndIf
\end{algorithmic}
\end{algorithm}

\subsection{WRA Updating Phase}
\label{WRA Updating Phase}
The node updates the WRA and $\overline{\lambda}$ according to~\eqref{eq1} and~\eqref{eq3} in the following two cases: when the node gets a transmission opportunity and is ready to initiate a reservation, or it reaches the updating period since the last time when the WRA was updated. The second case is used to identify whether it needs to access the channel using a smaller CW introduced in Section~\ref{Channel Access phase}.
\subsection{Reservation phase}
\label{Reservation phase}
If there is remaining real-time traffic in the queue in a transmission, the node is ready to initiate a reservation. The node first calculates the total time required for the next transmission as the reservation duration, which is based on the amount of real-time traffic loads in the queue and the corresponding control overhead. Then it calculates the $T_{offset}$ by the equation from~\eqref{eq1} to~\eqref{eq5} in turn. Then it compares with the recorded reservation table of other nodes to calculate the adjacent reservation time interval $T_{between}$, which represents the interval time between the beginning time of this reservation and the latest reservation time period of other nodes. If $T_{between}$ is less than $T_{frag}$, the transmission time fragmentation is considered to appear, and the adjacent reservation time interval needs to be corrected to the minimum reservation offset interval $T_{betmin}$. Finally, it generates the channel reservation period time $\tau (t_1,t_2)$ according to the reservation offset time and duration, making conflict-free panning with the existing reservation table of other nodes, and form RSI MPDU according to the above information and insert it into the AMPDU of this transmission.

\subsection{Channel Access phase}
\label{Channel Access phase}
Since nodes cannot broadcast WRA information when they have not accessed the channel, resulting in neighbor nodes use a smaller reservation offset time, making it still difficult for nodes to get a transmission opportunity to initiate a reservation. Therefore the protocol is additionally designed as follows.

When a node does not obtain a reservation opportunity and accesses the channel according to backoff, the used contention window is classified into three classes in order from smallest to largest: $CW_{low}$, $CW_{genal}$ and $CW_{high}$, and the $CW_{genal}$ is used to access the channel in the initial stage. If the $\overline{\lambda}$ exceeds the threshold $Th_{less}$, it adjusts its own combination of contention windows to $CW_{low}$ in order to compete for transmission opportunities; if it detects that all the links that need reservation have been given reservation opportunities, it will adjust to $CW_{high}$.
\subsection{Frame Format}
\label{Frame Format}
\begin{figure}[htbp]
    \centering
    \includegraphics[width=0.4\textwidth]{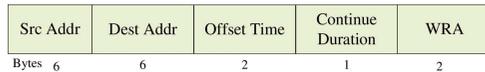}
    \caption{Frame format of reservation instruction field.}
	\label{fig3}
\end{figure}

RSI MPDU adopts the MAC header format of IEEE 802.11 protocol and indicates the type of RSI MSDU by the reserved value of the subtype field. In addition to the reservation owner source address, reservation link destination address, reservation offset time, and reservation duration included in the basic reservation mechanism, the WRA field with two bytes is added to the reservation instruction field for broadcasting the reservation capacity weight of this node information. Figure~\ref{fig3} shows the frame format of the reservation instruction field.

\section{SIMULATION RESULTS}
\label{SIMULATION RESULTS}
In this section, the performance of the proposed protocol is investigated. We use the simulation platform established based on \cite{simu1,simu2,simu3} and add the channel reservation access mechanism, using the basic parameters listed in Table~\ref{tab1} for configuration, where the routing layer uses the OLSR protocol, each simulation time is fifty seconds, and is repeated five times to take the average value. There are four scenarios, namely, short-hop topology, chain topology, cross-topology, and random topology. Three protocols are simulated for IEEE 802.11ax protocol, flexible reservation protocol SCRP proposed by Cheng \emph{et al.} \cite{13} and FTKN-CRM proposed in this paper, in terms of end-to-end throughput, delay, and energy consumption, respectively, where IEEE 802.11ax and SCRP are selected for simulation with the most suitable parameter configuration and use RTS/CTS mechanism to overcome hidden terminals.
\begin{table}[htbp]
\centering
\caption{Simulation Setup}
\begin{tabular}{cccc}\hline
Parameter & value \\\hline
Banwithch & 20Mhz\\
SIFS  & 64us \\
Length of slot  & 36us \\
Inertia factor & $0.6_{10}$\\
Transmission power & 24mW \\
Reception power & 13.5mW \\
Sleep power & 0.015mW  \\
Pacakge size & 1500Bytes\\
Real-time traffic $CW_{min}$ & 7\\
Real-time traffic $CW_{min}$ & 15\\
Maximum reservation duration & 5ms\\
$\alpha_{m}$ for local-traffic & 0.4\\
$\alpha_{m}$ for relay-traffic & 1\\
$Th_{buffer}$ & 40$\%$ \\\hline
\end{tabular}
\label{tab1}
\end{table}
\subsection{Short hop topology}
\label{Short hop topology}

\begin{figure}[htbp]
    \centering
    \includegraphics[width=0.42\textwidth]{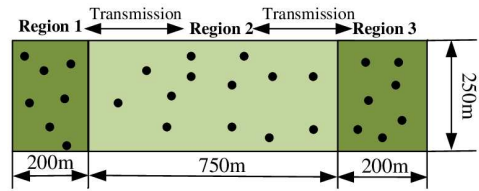}
    \caption{Short hop topology simulation scenario.}
	\label{fig4}
\end{figure}
\begin{figure}[htbp]
    \centering
    \includegraphics[width=0.42\textwidth]{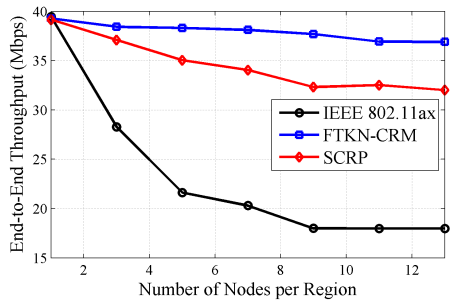}
    \caption{End-to-end throughput performance comparison in short-hop topology.}
	\label{fig5}
\end{figure}

The initial simulation establishes a simple topology with two-hop services and random placement of nodes shown in Figure~\ref{fig4}. The whole network is divided into three regions, where the nodes in region 1 and region 3 have the traffic with the destination for each other. The transmission range of the nodes is 750m, so the nodes at both ends communicate with each other through the node in the middle region to forward. Saturation traffic is generated for all nodes and the nodes are configured at random locations in the three regions according to the input number. The end-to-end throughput obtained from the simulation is shown in Figure~\ref{fig5}.
The simulation result shows that as the number of nodes increases, the SCAP can reduce collision by reservation, but the key nodes do not get enough opportunities to transmit, making the end-to-end throughput also drop significantly. While the FTKN-CRM protocol does not show a significant drop in throughput in the case of denser nodes. The throughput of the FTKN-CRM protocol is better than the two comparison schemes.
\subsection{Long-Chain Topology}
\label{Long-Chain Topology}
\begin{figure}[htbp]
    \centering
    \includegraphics[width=0.42\textwidth]{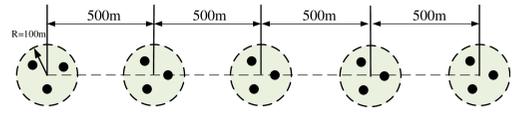}
    \caption{Long-chain topology simulation scenario.}
	\label{fig6}
\end{figure}
\begin{figure}[htbp]
    \centering
    \includegraphics[width=0.42\textwidth]{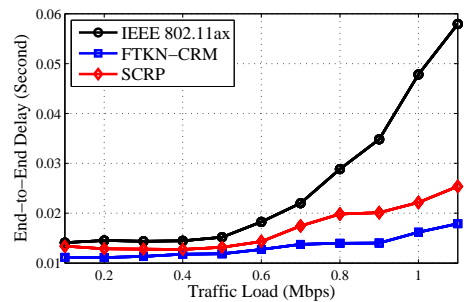}
    \caption{End-to-end throughput performance comparison in long-chain topology.}
	\label{fig7}
\end{figure}
\begin{figure}[htbp]
    \centering
    \includegraphics[width=0.42\textwidth]{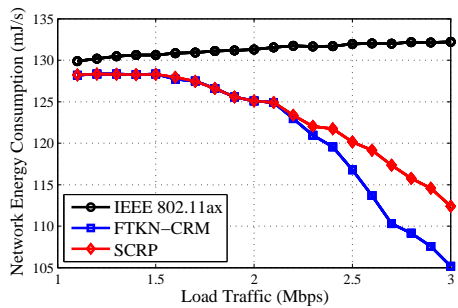}
    \caption{Energy consumption performance in long-chain topology.}
	\label{fig8}
\end{figure}

Four-hop chain typologies are established in figure 6. The distance between the centers of the two adjacent parts of the region is 500m, and the transmission range of the nodes is 750m. Three nodes are configured in each region, and the end-to-end delay and energy consumption simulation results are obtained as shown in Figure~\ref{fig7} and Figure~\ref{fig8}.

Simulation result for the end-to-end delay shows that all protocols have an increasing delay as the traffic load increases, the SCRP needs to ensure fairness in various traffic load cases, and thus the reservation offset time cannot be kept appropriate. For the comparison protocol, intermediate nodes in long chain topology do not get the priority transmission opportunity and thus cannot deliver the service to the peer node in time, although they need to forward more traffic and compete with more nodes. Therefore FTKN-CRM protocol has the lowest latency.

For the energy consumption performance, as the traffic loads of nodes gradually increase, the energy consumption of the IEEE 802.11ax protocol increases steadily and is much higher than the consumption under the reservation mechanism. It is because the nodes can sleep during the reservation time in the reservation mechanism, which can reduce the node energy consumption, so the energy consumption under the reservation mechanism gradually decreases as the percentage of reservation time increases. FTKN-CRM protocol flexibly adjusts the reservation offset time based on the neighbor nodes' WRA information, while SCRP protocol uses a fixed offset time so that the transmission opportunity is sometimes obtained through the competition when the channel is relatively free. Therefore FTKN-CRM protocol has less energy consumption than the SCRP protocol.
\subsection{Cross Topology}
\label{Cross Topology}
\begin{figure}[htbp]
    \centering
    \includegraphics[width=0.35\textwidth]{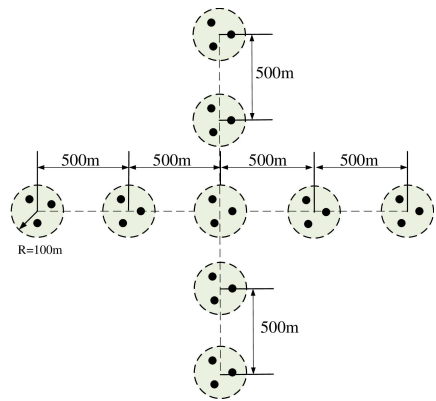}
    \caption{Cross topology simulation scenario.}
	\label{fig9}
\end{figure}
\begin{figure}[htbp]
    \centering
    \includegraphics[width=0.42\textwidth]{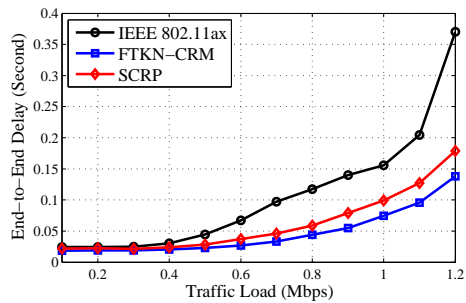}
    \caption{End-to-end delay simulation results in cross topology.}
	\label{fig10}
\end{figure}

The simulation establishes a total four-hop chain topology as shown in Figure~\ref{fig9}, with a distance of 500m between the centers of two adjacent parts of the region and a transmission range of 750m for the nodes, with three nodes configured in each region. Figure~\ref{fig10} shows the performance of end-to-end delay, it can be seen that as the offered load increases the end-to-end delay of IEEE 802.11ax protocol is significantly higher than the reservation mechanism, while the delay under FTKN-CRM protocol is lower than the end-to-end delay of SCRP protocol.
\subsection{Random Topology}
\label{Random Topology}
\begin{figure}[htbp]
    \centering
    \includegraphics[width=0.3\textwidth]{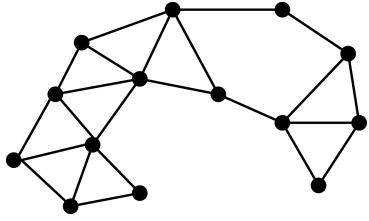}
    \caption{Generated random topology simulation scenario.}
	\label{fig11}
\end{figure}
A topology is randomly generated within $2000 \times 2000$ meters as shown in Figure~\ref{fig11}, which satisfies the requirement that all nodes can reach each other through relay nodes and limit the max distance between neighbor nodes, and two-hop and three-hop services are configured separately for simulation in this scenario.
\begin{figure}[htbp]
    \centering
    \includegraphics[width=0.42\textwidth]{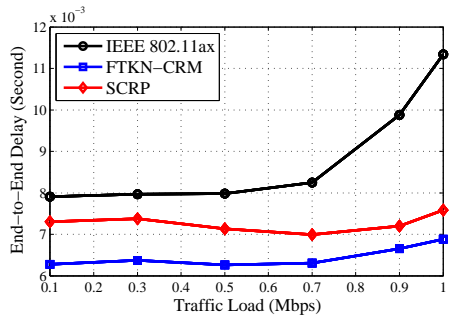}
    \caption{End-to-end delay configured with two-hop services in random topology.}
	\label{fig12}
\end{figure}
\begin{figure}[htbp]
    \centering
    \includegraphics[width=0.42\textwidth]{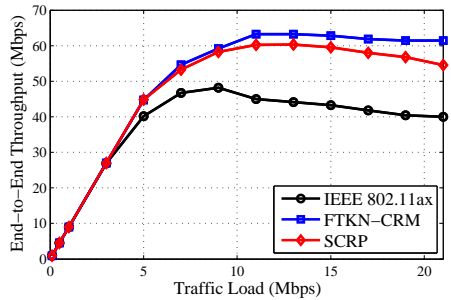}
    \caption{End-to-end throughput configured with two-hop services in random topology.}
	\label{fig13}
\end{figure}



First, the node is configured to randomly generate traffic with a destination of two hops away from this node for simulation, and the performance is shown in Figure~\ref{fig12} and Figure~\ref{fig13}. From the end-to-end throughput simulation results, it can be seen that as the load increases, the end-to-end throughput increases and then decreases under IEEE 802.11ax and SCRP protocols, while the performance of FTKN-CRM protocol increases and then remains almost unchanged and has a higher performance than IEEE 802.11ax and SCRP protocols. As for the end-to-end delay performance, the FTKN-CRM protocol has the lowest delay and the IEEE 802.11ax protocol has the highest delay.

\begin{figure}[htbp]
    \centering
    \includegraphics[width=0.42\textwidth]{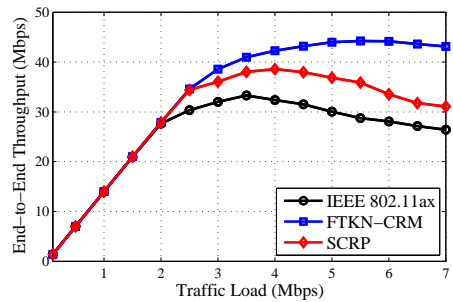}
    \caption{End-to-end performance configured with three-hop services in random topology.}
	\label{fig14}
\end{figure}

Finally, we increase the configured service type to a three-hop distance and get the simulation results shown in Figure~\ref{fig14}. At this point in the end-to-end throughput performance results, the end-to-end throughput of the IEEE 802.11ax and SCRP protocols show a more pronounced drop when the service is heavier, while the FTKN-CRM protocol tends to remain unchanged, indicating that the FTKN-CRM protocol can achieve end-to-end service transmission with reasonable use of channel resources even when the network is busier.

\section{Conclusion}
\label{Conclusion}
In order to improve the QoS in Ad Hoc networks, this paper proposes an adaptive channel reservation MAC protocol based on forwarding traffic of key nodes. Different reservation parameters are automatically configured for different nodes according to the forwarding service type and traffic loads adaptively, thus enabling key nodes to get enough transmission opportunities, and the end-to-end performance and energy consumption of the proposed protocol in Ad Hoc networks can be improved by simulation verification. The simple protocol and low signaling overhead of the algorithm in this paper have a strong application prospect and value in Ad Hoc networks.

\bmhead{Acknowledgments}
\label{ACKNOWLEDGEMENT}
This work was supported in part by the National Natural Science Foundations of CHINA (Grant No. 61871322, No. 61771392, and No. 61771390) and Science and Technology on Avionics Integration Laboratory and the Aeronautical Science Foundation of China (Grant No. 20185553035, and No. 201955053002).

\bmhead{Data Availability}
Data sharing not applicable to this article as no datasets were generated or analysed during the current study.


\bibliographystyle{gbt7714-numerical}
\bibliography{sn-bibliography}



\end{document}